\documentclass[pra,10pt,showpacs,amsmath,twocolumn,floatfix,eqsecnum]{revtex4}
\usepackage{amsmath}
\usepackage{amsfonts}
\usepackage{graphicx}

\newcommand{\real}{\operatorname{Re}}

\newcommand{\parti}[2]{\frac{\partial #1}{\partial #2}}
\newcommand{\partit}[2]{\frac{\partial^2 #1}{\partial #2^2}}

\newcommand{\ket}[1]{|#1\rangle}
\newcommand{\Ket}[1]{\left|#1\right\rangle}
\newcommand{\bra}[1]{\langle#1|}
\newcommand{\Bra}[1]{\left\langle#1\right|}

\newcommand{\avg}[1]{\langle#1\rangle}
\newcommand{\Avg}[1]{\left\langle#1\right\rangle}

\newcommand{\bk}[1]{\left(#1\right)}
\newcommand{\Bk}[1]{\left[#1\right]}
\newcommand{\BK}[1]{\left\{#1\right\}}

\newcommand{\trace}{\operatorname{tr}}

\begin{document}
\title{Optimal waveform estimation for classical and quantum systems
via time-symmetric smoothing}

\author{Mankei Tsang}

\email{mankei@mit.edu}

\affiliation{Research Laboratory of Electronics,
Massachusetts Institute of Technology, Cambridge, Massachusetts
02139, USA}






\date{\today}

\begin{abstract}
  Classical and quantum theories of time-symmetric smoothing, which
  can be used to optimally estimate waveforms in classical and quantum
  systems, are derived using a discrete-time approach, and the
  similarities between the two theories are emphasized. Application of
  the quantum theory to homodyne phase-locked loop design for phase
  estimation with narrowband squeezed optical beams is studied. The
  relation between the proposed theory and Aharonov \textit{et al.}'s
  weak value theory is also explored.
\end{abstract}
\pacs{03.65.Ta, 03.65.Yz, 42.50.Dv}

\maketitle
\section{Introduction}
\begin{figure}[htbp]
\centerline{\includegraphics[width=0.48\textwidth]{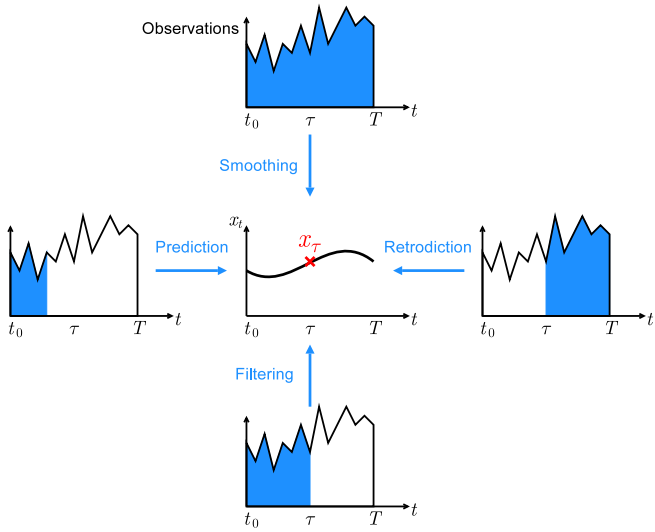}}
\caption{(Color online). Four classes of estimation problems, depending
 on the observation time interval relative to $\tau$, the time at which
 the signal is to be estimated.}
\label{classes}
\end{figure}

Estimation theory is the science of determining the state of a system,
such as a dice, an aircraft, or the weather in Boston, from noisy
observations \cite{jazwinski,vantrees,crassidis,simon}. As shown in
Fig.~\ref{classes}, estimation problems can be classified into four
classes, namely, prediction, filtering, retrodiction, and smoothing.
For applications that do not require real-time data, such as sensing
and communication, smoothing is the most accurate estimation
technique.

I have recently proposed a time-symmetric quantum theory of smoothing,
which allows one to optimally estimate classical diffusive Markov
random processes, such as gravitational waves or magnetic fields,
coupled to a quantum system, such as a quantum mechanical oscillator
or an atomic spin ensemble, under continuous measurements
\cite{tsang_smooth}.  In this paper, I shall demonstrate in more
detail the derivation of this theory using a discrete-time approach,
and how it closely parallels the classical time-symmetric smoothing
theory proposed by Pardoux \cite{pardoux}.  I shall apply the
theory to the design of homodyne phase-locked loops (PLL) for
narrowband squeezed optical beams, as previously considered by Berry
and Wiseman \cite{berry}. I shall show that their approach can be
regarded as a special case of my theory, and discuss how their results
can be generalized and improved.  I shall also discuss the weak value
theory proposed by Aharonov \textit{et al.}\ \cite{aav} in relation
with the smoothing theory, and how their theory may be regarded as a
smoothing theory for quantum degrees of freedom. In particular, the
smoothing quasiprobability distribution proposed in
Ref.~\cite{tsang_smooth} is shown to naturally arise from the
statistics of weak position and momentum measurements.

This paper is organized as follows: In Sec.~\ref{classical}, Pardoux's
classical time-symmetric smoothing theory is derived using a
discrete-time approach, which is then generalized to the quantum
regime for hybrid classical-quantum smoothing in
Sec.~\ref{hybrid}. Application of the hybrid classical-quantum
smoothing theory to PLL design is studied in Sec.~\ref{adaptive}. The
relation between the smoothing theory and Aharonov \textit{et al.}'s
weak value theory is then discussed in
Sec.~\ref{weak}. Sec.~\ref{conclusion} concludes the paper and points
out some possible extensions of the proposed theory.

\section{\label{classical}Classical smoothing}
\subsection{Problem statement}
\begin{figure}[htbp]
\centerline{\includegraphics[width=0.4\textwidth]{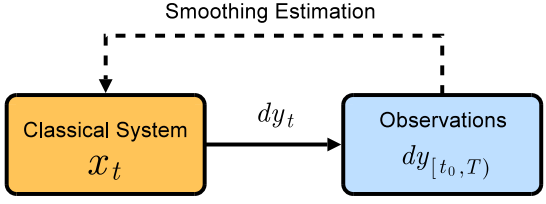}}
\caption{(Color online). The classical smoothing problem.}
\label{classical_smoothing}
\end{figure}
Consider the classical smoothing problem depicted in
Fig.~\ref{classical_smoothing}.  Let
\begin{align}
x_t &\equiv \Bk{\begin{array}{c}x_{1t}\\ x_{2t}\\\vdots\\ x_{nt}\end{array}}
\end{align}
be a vectoral diffusive Markov random process that satisfies the
system It\=o differential equation \cite{jazwinski}
\begin{align}
dx_t &= A(x_t,t)dt + B(x_t,t)dW_t,
\label{system}
\end{align}
where $dW_t$ is a vectoral Wiener increment with mean and covariance
matrix given by
\begin{align}
\Avg{dW_t} &= 0,
\\
\Avg{dW_t dW_t^T} &= Q(t)dt.
\end{align}
The superscript $^T$ denotes
the transpose. The vectoral observation process $dy_t$ satisfies the
observation It\=o equation
\begin{align}
dy_t &= C(x_t,t)dt + dV_t,
\label{observ}
\end{align}
where $dV_t$ is another vectoral Wiener increment with mean and
covariance matrix given by
\begin{align}
\Avg{dV_t} &= 0,
\\
\Avg{dV_tdV_t^T} &= R(t)dt.
\end{align}
For generality and later purpose, $dW_t$ and $dV_t$ are assumed to be
correlated, with covariance
\begin{align}
\Avg{dW_tdV_t^T} &= S(t)dt.
\end{align}
Define the observation record in the time interval $[t_1,t_2)$ as
\begin{align}
dy_{[t_1,t_2)} &\equiv \BK{dy_t, t_1\le t < t_2}.
\end{align}
The goal of smoothing is to calculate the conditional probability
density of $x_\tau$, given the observation record $dy_{[t_0,T)}$ in
the time interval $t_0\le \tau \le T$.

It is more intuitive to consider the problem in discrete time
first. The discrete-time system and observation equations
(\ref{system}) and (\ref{observ}) are
\begin{align}
\delta x_t &= A(x_t,t)\delta t + B(x_t,t)\delta W_t,
\label{discrete_system}\\
\delta y_t &= C(x_t,t)\delta t + \delta V_t.
\label{discrete_observ}
\end{align}
The observation record
\begin{align}
\delta y_{[t_0,T-\delta t]} &\equiv 
\BK{\delta y_{t_0},\delta y_{t_0+\delta t},\dots,\delta y_{T-\delta t}}
\end{align}
also becomes discrete.  The covariance matrices for the
increments are
\begin{align}
\Avg{\delta W_t\delta W_t^T} &= Q(t)\delta t,
\\
\Avg{\delta V_t\delta V_t^T} &= R(t)\delta t,
\\
\Avg{\delta W_t\delta V_t^T} &= S(t)\delta t,
\end{align}
and the increments at different times are independent of one
another.  Because $\delta W_t$ and $\delta V_t$ are proportional to
$\sqrt{\delta t}$, one should keep all linear \emph{and} quadratic
terms of the Wiener increments in an equation according to It\=o
calculus when taking the continuous time limit.

With correlated $\delta W_t$ and $\delta V_t$, it is preferable, for
technical reasons, to rewrite the system equation
(\ref{discrete_system}) as \cite{crassidis}
\begin{align}
\delta x_t &= A(x_t,t)\delta t+B(x_t,t) \delta W_t
\nonumber\\&\quad
+D(x_t,t)\Bk{\delta y_t-C(x_t,t)\delta t - \delta V_t},
\end{align}
where $D(x_t,t)$ can be arbitrarily set because the expression in
square brackets is zero. The system equation becomes
\begin{align}
\delta x_t &= \Bk{A(x_t,t)-D(x_t,t)C(x_t,t)}\delta t+
D(x_t,t)\delta y_t 
\nonumber\\&\quad
+B(x_t,t)\delta W_t - D(x_t,t)\delta V_t.
\end{align}
The new system noise is
\begin{align}
\delta Z_t &\equiv B(x_t,t)\delta W_t - D(x_t,t)\delta V_t,
\\
\Avg{\delta Z_t\delta Z_t^T} &= 
\big[B(x_t,t)Q(t)B^T(x_t,t)
\nonumber\\&\quad
+D(x_t,t)R(t)D^T(x_t,t)
\nonumber\\&\quad
-B(x_t,t)S(t)D^T(x_t,t)
\nonumber\\&\quad
-D(x_t,t)S^T(t)B^T(x_t,t)\big]\delta t.
\end{align}
The covariance between the new system noise $\delta Z_t$ and
the observation noise $\delta V_t$ is
\begin{align}
\Avg{\delta Z_t\delta V_t^T} &= 
\Bk{B(x_t,t)S(t) - D(x_t,t)R(t)}\delta t,
\end{align}
and can be made to vanish if one lets
\begin{align}
D(x_t,t) &= B(x_t,t)S(t)R^{-1}(t).
\end{align}
The new equivalent system and observation model is then
\begin{align}
\delta x_t &= A(x_t,t)\delta t 
+B(x_t,t)S(t)R^{-1}(t)\Bk{\delta y_t -C(x_t,t)\delta t}
\nonumber\\&\quad
+B(x_t,t)\delta U_t,
\label{new_system}\\
\delta y_t &= C(x_t,t)\delta t + \delta V_t,
\label{new_observ}
\end{align}
with covariances
\begin{align}
\Avg{\delta U_t\delta U_t^T} &= \Bk{Q(t)-S(t)R^{-1}S^T(t)}\delta t,
\\
\Avg{\delta V_t\delta V_t^T} &= R(t)\delta t,
\\
\Avg{\delta U_t\delta V_t^T} &= 0.
\end{align}
The new system and observation noises are now independent, but note
that $\delta x_t$ becomes dependent on $\delta y_t$.

\subsection{Time-symmetric approach}
According to the Bayes theorem, the smoothing probability density for
$x_\tau$ can be expressed as
\begin{align}
P(x_\tau|\delta y_{[t_0,T-\delta t]}) &= 
\frac{P(\delta y_{[t_0,T-\delta t]}|x_\tau) P(x_\tau)}
{P(\delta y_{[t_0,T-\delta t]})},
\label{bayes}
\\
P(\delta y_{[t_0,T-\delta t]}) &=
\int dx_\tau P(\delta y_{[t_0,T-\delta t]}|x_\tau) P(x_\tau),
\end{align}
where
\begin{align}
\int dx_\tau &\equiv \int dx_{1\tau}\dots \int dx_{n\tau}
\end{align}
and $P(x_\tau)$ is the \textit{a priori} probability density, which
represents one's knowledge of $x_\tau$ absent any observation.
Functions of $x_\tau$ are assumed to also depend implicitly on $\tau$.
Splitting $\delta y_{[t_0,T-\delta t]}$ into the past record
\begin{align}
\delta y_{\textrm{past}} &\equiv \delta y_{[t_0,\tau-\delta t]}
\end{align}
and the future record 
\begin{align}
\delta y_{\textrm{future}}&\equiv \delta y_{[\tau,T-\delta t]}
\end{align}
relative to time $\tau$, $P(\delta y_{[t_0,T)}|x_\tau)$ in
Eq.~(\ref{bayes}) can be rewritten as
\begin{align}
P(\delta y_{[t_0,T-\delta t]}|x_\tau)
&= P(\delta y_{\textrm{past}},\delta y_{\textrm{future}}|x_\tau)
\nonumber\\
&=P(\delta y_{\textrm{future}}|\delta y_{\textrm{past}},x_\tau)
P(\delta y_{\textrm{past}}|x_\tau).
\end{align}
Because $\delta V_t$ are independent increments, the future record is
independent of the past record given $x_\tau$, and
\begin{align}
P(\delta y_{[t_0,T-\delta t]}|x_\tau)
&=P(\delta y_{\textrm{future}}|x_\tau)P(\delta y_{\textrm{past}}|x_\tau).
\end{align}
Equation (\ref{bayes}) becomes
\begin{align}
P(x_\tau|\delta y_{[t_0,T-\delta t]})
&= \frac{P(\delta y_{\textrm{future}}|x_\tau)P(\delta y_{\textrm{past}}|x_\tau)P(x_\tau)}
{\int dx_\tau (\textrm{numerator})}
\nonumber\\
&=
\frac{P(\delta y_{\textrm{future}}|x_\tau)P(x_\tau|\delta y_{\textrm{past}})}
{\int dx_\tau (\textrm{numerator})}.
\label{twofilter}
\end{align}
Thus, the smoothing density can be obtained by combining the
filtering probability density $P(x_\tau|\delta y_{\textrm{past}})$ and
a retrodictive likelihood function $P(\delta
y_{\textrm{future}}|x_\tau)$.

\subsection{Filtering}
To derive an equation for the filtering probability density
$P(x_\tau|\delta y_{\textrm{past}})$, first express $P(x_{t+\delta
  t}|\delta y_{[t_0,t]})$ in terms of $P(x_{t}|\delta y_{[t_0,t]})$ as
\begin{align}
P(x_{t+\delta t}|\delta y_{[t_0,t]})
&=
\int dx_t P(x_{t+\delta t},x_t|\delta y_{[t_0,t]})
\nonumber\\
&=
\int dx_t P(x_{t+\delta t}|x_t,\delta y_{[t_0,t]})
P(x_t|\delta y_{[t_0,t]}).
\end{align}
$P(x_{t+\delta t}|x_t,\delta y_{[t_0,t]})= P(x_{t+\delta t}|x_t,\delta
y_t,\delta y_{[t_0,t-\delta t]})$ can be determined from the system
equation (\ref{new_system}) and is equal to $P(x_{t+\delta
  t}|x_t,\delta y_{t})$, due to the Markovian nature of the system
process. So
\begin{align}
P(x_{t+\delta t}|\delta y_{[t_0,t]})
&=
\int dx_t P(x_{t+\delta t}|x_t,\delta y_{t})
P(x_t|\delta y_{[t_0,t]}),
\label{ck}
\end{align}
which is a generalized Chapman-Kolmogorov equation \cite{gardiner}.
$P(x_{t+\delta t}|x_t,\delta y_{t})$ is
\begin{align}
&P(x_{t+\delta t}|x_t,\delta y_{t})
\propto
\nonumber\\&\quad
\exp\bigg\{-\frac{1}{2}\delta Z_t^T
\Bk{B(x_t,t)Q(t)B^T(x_t,t)\delta t}^{-1}
\delta Z_t\bigg\},
\label{gauss_system}
\end{align}
where
\begin{align}
\delta Z_t &\equiv x_{t+\delta t}-x_t
-A(x_t,t)\delta t
\nonumber\\&\quad
+B(x_t,t)S(t)R^{-1}(t)
\Bk{\delta y_t - C(x_t,t)\delta t}.
\end{align}
Next, write $P(x_t|\delta y_{[t_0,t]})$ in terms of $P(x_t|\delta
y_{[t_0,t-\delta t]})$ using the Bayes theorem as
\begin{align}
P(x_t|\delta y_{[t_0,t]})
&= P(x_t|\delta y_{[t_0,t-\delta t]},\delta y_t)
\nonumber\\
&=\frac{P(\delta y_t|x_t,\delta y_{[t_0,t-\delta t]})P(x_t|\delta y_{[t_0,t-\delta t]})}
{\int dx_t (\textrm{numerator})}
\nonumber\\
&=
\frac{P(\delta y_t|x_t)P(x_t|\delta y_{[t_0,t-\delta t]})}
{\int dx_t
(\textrm{numerator})},
\label{bayes2}
\end{align}
where $P(\delta y_t|x_t,\delta y_{[t_0,t-\delta t]}) = P(\delta
y_t|x_t)$ due to the Markovian property of the observation
process. $P(\delta y_t|x_t)$ is determined by the observation equation
(\ref{new_observ}) and given by
\begin{align}
P(\delta y_t|x_t)
&\propto
\exp\bigg\{-\frac{1}{2}\Bk{\delta y_t-C(x_t,t)\delta t}^T
\Bk{R(t)\delta t}^{-1}\nonumber\\&\quad\times
\Bk{\delta y_t-C(x_t,t)\delta t}\bigg\}.
\label{gauss_observ}
\end{align}
Hence, starting with the \textit{a priori} probability
density $P(x_{t_0})$, one can solve for $P(x_\tau|\delta y_{\textrm{past}})$
by iterating the formula
\begin{align}
P(x_{t+\delta t}|\delta y_{[t_0,t]})
&=\int dx_t P(x_{t+\delta t}|x_t,\delta y_t)
\nonumber\\&\quad\times
\frac{P(\delta y_t|x_t)P(x_t|\delta y_{[t_0,t-\delta t]})}
{\int dx_t (\textrm{numerator})}.
\label{iterate}
\end{align}
To obtain a stochastic differential equation for the filtering
probability density, defined as
\begin{align}
F(x,t) &\equiv P(x_t=x|dy_{[t_0,t)})
\end{align}
in the continuous time limit, one should expand Eq.~(\ref{iterate}) to
first order with respect to $\delta t$ and second order with respect
to $\delta y_t$ in a Taylor series, then apply the rules of It\=o
calculus. The result is the Kushner-Stratonovich (KS) equation
\cite{jazwinski,kushner}, generalized for correlated system and
observation noises by Fujisaki \textit{et al.}\ \cite{fujisaki},
given by
\begin{align}
dF &= -dt\sum_\mu \parti{}{x_\mu}\bk{A_\mu F}
\nonumber\\&\quad
+\frac{dt}{2}\sum_{\mu,\nu}\parti{^2}{x_\mu\partial x_\nu}
\Bk{\bk{BQB^T}_{\mu\nu}F}
\nonumber\\&\quad
+\bk{C-\avg{C}_F}^TR^{-1}d\eta_t F
\nonumber\\&\quad
-\sum_{\mu}\parti{}{x_\mu}
\Bk{\bk{BSR^{-1}d\eta_t}_\mu F},
\label{kushner}
\end{align}
where
\begin{align}
dF &\equiv F(x,t+dt)-F(x,t),
\\
\avg{C}_F &\equiv \int dx C(x,t)F(x,t),
\\
d\eta_t &\equiv dy_t - dt\avg{C}_F.
\end{align}
The initial condition is
\begin{align}
F(x,t_0) &= P(x_{t_0}).
\end{align}
$d\eta_t$ is called the innovation process and is also a Wiener
increment with covariance matrix $R(t)dt$ \cite{frost,fujisaki}.

A linear stochastic equation for an unnormalized $F$ is called the
Duncan-Mortensen-Zakai (DMZ) equation \cite{pardoux,zakai}, given by
\begin{align}
df &= -dt\sum_\mu \parti{}{x_\mu}\bk{A_\mu f}
\nonumber\\&\quad
+\frac{dt}{2}\sum_{\mu,\nu}\parti{^2}{x_\mu\partial x_\nu}
\Bk{\bk{BQB^T}_{\mu\nu}f}
\nonumber\\&\quad
+C^TR^{-1}dy_t f
-\sum_{\mu}\parti{}{x_\mu}
\Bk{\bk{BSR^{-1}dy_t}_\mu f},
\label{zakai}
\end{align}
where the normalization is
\begin{align}
F(x,t) &= \frac{f(x,t)}{\int dx f(x,t)}.
\end{align}

\subsection{\label{retro_smooth}Retrodiction and smoothing}
To solve for the retrodictive likelihood function $P(\delta
y_{\textrm{future}}|x_\tau)$, note that
\begin{align}
P(\delta y_{\textrm{future}}) &=
\int dx_\tau P(\delta y_{\textrm{future}}|x_\tau)P(x_\tau),
\label{retro}
\end{align}
but $P(\delta y_{\textrm{future}})$ can also be expressed 
in terms of the multitime probability density as
\begin{align}
P(\delta y_{[\tau,T-\delta t]})
&= \int Dx_{[\tau,T]}
P(x_{[\tau,T]},\delta y_{[\tau,T-\delta t]}),
\end{align}
where
\begin{align}
x_{[\tau,T]} &\equiv \BK{x_\tau,x_{\tau+\delta t},\dots,x_T},
\\
\int Dx_{[\tau,T]} &\equiv \int dx_\tau \int dx_{\tau+\delta t}\dots\int dx_T.
\end{align}
The multitime density can be rewritten as
\begin{align}
P(x_{[\tau,T]},\delta y_{[\tau,T-\delta t]})
&=
P(x_T|x_{[\tau,T-\delta t]},\delta y_{[\tau,T-\delta t]})
\nonumber\\&\quad\times
P(x_{[\tau,T-\delta t]},\delta y_{[\tau,T-\delta t]}).
\label{multitime}
\end{align}
Again using the Markovian property of the system process,
\begin{align}
P(x_T|x_{[\tau,T-\delta t]},\delta y_{[\tau,T-\delta t]}) =P(x_T|x_{T-\delta
  t},\delta y_{T-\delta t}),
\label{markov_system}
\end{align}
which can be determined from the system equation (\ref{new_system})
and is given by Eq.~(\ref{gauss_system}). Furthermore,
$P(x_{[\tau,T-\delta t]},\delta y_{[\tau,T-\delta t]})$ in Eq.~(\ref{multitime})
can be expressed as
\begin{align}
P(x_{[\tau,T-\delta t]},\delta y_{[\tau,T-\delta t]})
&=
P(\delta y_{T-\delta t}|x_{[\tau,T-\delta t]},\delta y_{[\tau,T-2\delta t]})
\nonumber\\&\quad\times
P(x_{[\tau,T-\delta t]},\delta y_{[\tau,T-2\delta t]}).
\label{multitime2}
\end{align}
Using the Markovian property of the observation process,
\begin{align}
P(\delta y_{T-\delta t}|x_{[\tau,T-\delta t]},\delta
y_{[\tau,T-2\delta t]}) = P(\delta y_{T-\delta t}|x_{T-\delta t}),
\label{markov_observe}
\end{align}
which can be determined from the observation equation
(\ref{new_observ}) and is given by Eq.~(\ref{gauss_observ}). Applying
Eqs.~(\ref{multitime}), (\ref{markov_system}), (\ref{multitime2}), and
(\ref{markov_observe}) repeatedly, one obtains
\begin{align}
P(\delta y_{[\tau,T-\delta t]})
&= \int dx_T \int dx_{T-\delta t} P(x_T|x_{T-\delta t},\delta y_{T-\delta t})
\nonumber\\&\quad\times
 P(\delta y_{T-\delta t}|x_{T-\delta t})
\nonumber\\&\quad\times
\int dx_{T-2\delta t}
P(x_{T-\delta t}|x_{T-2\delta t},\delta y_{T-2\delta t})
\nonumber\\&\quad\times
 P(\delta y_{T-2\delta t}|x_{T-2\delta t})\dots
\nonumber\\&\quad\times
\int dx_\tau P(x_{\tau+\delta t}|x_\tau,\delta y_\tau)
\nonumber\\&\quad\times
P(\delta y_{\tau}|x_\tau)P(x_\tau).
\label{expand}
\end{align}
Comparing this equation with Eq.~(\ref{retro}), $P(\delta
y_{\textrm{future}}|x_\tau)$ can be expressed as
\begin{align}
P(\delta y_{\textrm{future}}|x_\tau)
&= 
P(\delta y_{\tau}|x_\tau)
\nonumber\\&\quad\times
\int dx_{\tau+\delta t} P(x_{\tau+\delta t}|x_\tau,\delta y_\tau)\dots
\nonumber\\&\quad
\times  P(\delta y_{T-2\delta t}|x_{T-2\delta t})
\nonumber\\&\quad\times
\int dx_{T-\delta t}
P(x_{T-\delta t}|x_{T-2\delta t},\delta y_{T-2\delta t})
\nonumber\\&\quad\times
P(\delta y_{T-\delta t}|x_{T-\delta t})
\nonumber\\&\quad\times
\int dx_T P(x_T|x_{T-\delta t},\delta y_{T-\delta t}).
\label{retro2}
\end{align}
Defining the unnormalized retrodictive likelihood function at time $t$
as
\begin{align}
g(x,t) \propto P(dy_{[t,T)}|x_t=x),
\end{align}
one can derive a linear backward stochastic differential equation for
$g(x,t)$ by applying It\=o calculus backward in time to
Eq.~(\ref{retro2}). The result is \cite{pardoux}
\begin{align}
-dg &= dt\sum_\mu A_\mu\parti{g}{x_\mu} 
+\frac{dt}{2}\sum_{\mu,\nu}\bk{BQB^T}_{\mu\nu}\parti{^2g}{x_\mu\partial x_\nu}
\nonumber\\&\quad
+C^TR^{-1}dy_t g
+\sum_{\mu}\bk{BSR^{-1}dy_t}_\mu\parti{g}{x_\mu}.
\label{backward_zakai}
\end{align}
which is the adjoint equation of the forward DMZ equation
(\ref{zakai}), to be solved backward in time in the backward It\=o
sense, defined by
\begin{align}
-dg &\equiv g(x,t-dt)-g(x,t),
\end{align}
with the final condition
\begin{align}
g(x,T) \propto 1.
\end{align}
The adjoint equation with respect to a linear differential equation
\begin{align}
df(x,t) &= \hat L f(x,t)
\end{align}
is defined as
\begin{align}
-dg(x,t) &= \hat L^\dagger g(x,t),
\end{align}
where $\hat L$ is a linear operator and $\hat L^\dagger$ is the
adjoint of $\hat L$, defined by
\begin{align}
\Avg{g(x),\hat L f(x)}
&= 
\Avg{\hat L^\dagger g(x), f(x)}
\end{align}
with respect to the inner product
\begin{align}
\Avg{g(x), f(x)} &\equiv \int dx g(x)f(x).
\end{align}

After solving Eq.~(\ref{zakai}) for $f(x,\tau)$ and
Eq.~(\ref{backward_zakai}) for $g(x,\tau)$, the smoothing probability
density is
\begin{align}
h(x,\tau) \equiv
P(x_\tau=x|dy_{[t_0,T)}) &= 
\frac{g(x,\tau)f(x,\tau)}{\int dx g(x,\tau)f(x,\tau)}.
\label{smooth}
\end{align}
Since $f(x,\tau)$ and $g(x,\tau)$ are solutions of adjoint equations,
their inner product, which appears as the denominator of
Eq.~(\ref{smooth}), is constant in time \cite{pardoux}.  The
denominator also ensures that $h(x,\tau)$ is normalized, and
$f(x,\tau)$ and $g(x,\tau)$ need not be normalized separately.

The estimation errors depend crucially on the statistics of $x_t$.  If
any component of $x_t$, say $x_{\mu t}$, is constant in time, then
filtering of that particular component is as accurate as smoothing,
for the simple reason that $P(x_{\mu\tau}|dy_{[t_0,T)})$ must be the
same for any $\tau$, and one can simply estimate $x_{\mu\tau}$ at the
end of the observation interval ($\tau = T$) using filtering
alone. This also means that smoothing is not needed when one only
needs to detect the presence of a signal in detection problems
\cite{vantrees}, since the presence can be regarded as a constant
binary parameter within a certain time interval.  In general, however,
smoothing can be significantly more accurate than filtering for the
estimation of a fluctuating random process in the middle of the
observation interval. Another reason for modeling unknown signals as
random processes is robustness, as introducing fictitious system noise
can improve the estimation accuracy when there are modeling errors
\cite{jazwinski,simon}.

\subsection{\label{mfp} Linear time-symmetric smoothing}
If $f$, $g$, and $h$ are Gaussian, one can just solve for their means
and covariance matrices, which completely determine the probability
densities. This is the case when the \textit{a priori} probability
density $P(x_{t_0})$ is Gaussian, and
\begin{align}
A(x_t,t) &= J(t)x_t,
\label{linearA}\\
B(x_t,t) &= B(t),
\label{linearB}\\
C(x_t,t) &= K(t)x_t.
\label{linearC}
\end{align}
The means and covariance matrices of $f$, $g$, and $h$ can then be
solved using the linear Mayne-Fraser-Potter (MFP) smoother
\cite{mayne}. The smoother first solves for the mean $x'$ and
covariance matrix $\Sigma$ of $f$ using the Kalman filter
\cite{jazwinski}, given by
\begin{align}
dx' &=  J x'dt + \Gamma\bk{dy-Kx'dt},
\\
\Gamma &\equiv \bk{\Sigma  K^T+BS}R^{-1},
\\
d\Sigma  &= \bk{J\Sigma  + \Sigma J^T - \Gamma  R\Gamma ^T + BQB^T}dt,
\end{align}
with the initial conditions at $t_0$ determined from $P(x_{t_0})$.
The mean $x''$ and covariance matrix $\Xi$ of $g$ are then solved
using a backward Kalman filter,
\begin{align}
-dx'' &= -Jx''dt + \Upsilon(dy-Kx''dt),
\\
\Upsilon &\equiv \bk{\Xi K^T+BS}R^{-1},
\\
-d\Xi &= \bk{-J\Xi-\Xi J^T - \Upsilon R\Upsilon^T + BQB^T}dt,
\end{align}
with the final condition $\Xi_T^{-1}x''_T = 0$ and $\Xi_T^{-1} =
0$. In practice, the information filter formalism should be used to
solve the backward filter, in order to avoid dealing with the infinite
covariance matrix at $T$ \cite{crassidis,mayne}. Finally, the
smoothing mean $\tilde x_\tau$ and covariance matrix $\Pi_\tau$ are
\begin{align}
\tilde x_\tau &= \Pi_\tau\bk{\Sigma _\tau^{-1}x'_\tau+\Xi_{\tau}^{-1}x''_\tau},
\\
\Pi_{\tau} &= \bk{\Sigma _\tau^{-1}+\Xi_{\tau}^{-1}}^{-1}.
\end{align}
Note that $x''$ and $\Xi$ are the mean and covariance matrix of a
likelihood function $P(dy_{[t,T)}|x_t)$ and not those of a conditional
probability density $P(x_t|dy_{[t,T)})$, so to perform optimal
retrodiction ($\tau = t_0$) one should still combine $x''$ and $\Xi$
with the \textit{a priori} values \cite{wall}.

\section{\label{hybrid}Hybrid classical-quantum smoothing}
\subsection{Problem statement}
\begin{figure}[htbp]
\centerline{\includegraphics[width=0.48\textwidth]{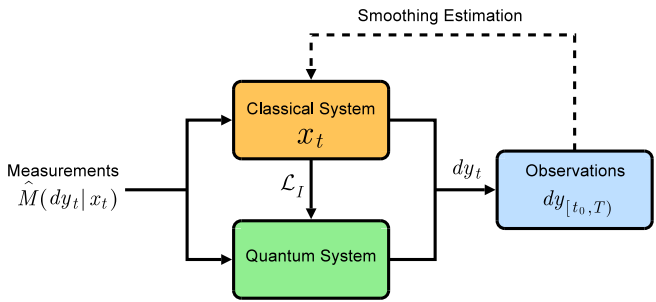}}
\caption{(Color online). Schematic of the hybrid classical-quantum
  smoothing problem.}
\label{hybrid_smooth}
\end{figure}

Consider the problem of waveform estimation in a hybrid
classical-quantum system depicted in Fig.~\ref{hybrid_smooth}. The
classical system produces a vectoral classical diffusive Markov random
process $x_t$, which obeys Eq.~(\ref{system}) and is coupled to the
quantum system.  The goal is to estimate $x_\tau$ via continuous
measurements of both systems. This setup is slightly more general than
that considered in \cite{tsang_smooth}; here the observations can
also depend on $x_t$. This allows one to apply the theory to PLL
design for squeezed beams, as considered by Berry and Wiseman
\cite{berry}, and potentially to other quantum estimation problems as
well \cite{was}. The statistics of $x_t$ are assumed to be unperturbed
by the coupling to the quantum system, in order to avoid the
nontrivial issue of quantum backaction on classical systems
\cite{backaction}. For simplicity, in this section we neglect the
possibility that the system noise driving the classical system is
correlated with the observation noise, although the noise driving the
quantum system can still be correlated with the observation noise due
to quantum measurement backaction.  Just as in the classical smoothing
problem, the hybrid smoothing problem is solved by calculating the
smoothing probability density $P(x_\tau|dy_{[t_0,T)})$.

\subsection{Time-symmetric approach}
Because a quantum system is involved, one may be tempted to use a
hybrid density operator \cite{backaction,tsang_smooth,berry,was} to
represent one's knowledge about the hybrid classical-quantum
system. The hybrid density operator $\hat\rho(x_\tau)$ describes the
joint classical and quantum statistics of a hybrid system, with the
marginal classical probability density for $x_\tau$ and the marginal
density operator for the quantum system given by
\begin{align}
P(x_\tau) &= \trace\Bk{\hat\rho(x_\tau)},
\\
\hat\rho(\tau) &= \int dx_\tau \hat\rho(x_\tau),
\end{align}
respectively.  The hybrid operator can also be regarded as a special
case of the quantum density operator, when certain degrees of freedom
are approximated as classical. Unfortunately, the density operator in
conventional predictive quantum theory can only be conditioned upon
past observations and not future ones, so it cannot be used as a
quantum version of the smoothing probability density.

The classical time-symmetric smoothing theory, as a combination of
prediction and retrodiction, offers an important clue to how one can
circumvent the difficulty of defining the smoothing quantum state.
Again casting the problem in discrete time, and defining a hybrid
effect operator as $\hat E(\delta y_{\textrm{future}}|x_\tau)$, which
can be used to determine the statistics of future observations given a
density operator at $\tau$,
\begin{align}
P(\delta y_{\textrm{future}}) &= 
\int dx_\tau 
\trace\Bk{\hat E(\delta y_{\textrm{future}}|x_\tau)\hat\rho(x_\tau)},
\end{align}
one may write, in analogy with Eq.~(\ref{twofilter})
\cite{tsang_smooth},
\begin{align}
P(x_\tau|\delta y_{[t_0,T-\delta t]})
&=
\frac{P(x_\tau,\delta y_{\textrm{future}}|\delta y_{\textrm{past}})}
{P(\delta y_{\textrm{future}}|\delta y_{\textrm{past}})}
\nonumber\\
&=
\frac{\trace[\hat E(\delta y_{\textrm{future}}|x_\tau)
\hat\rho(x_\tau|\delta y_{\textrm{past}})]}
{\int dx_\tau \trace[\hat E(\delta y_{\textrm{future}}|x_\tau)
\hat\rho(x_\tau|\delta y_{\textrm{past}})]},
\label{two_qfilter}
\end{align}
where $\hat\rho(x_\tau|\delta y_{\textrm{past}})$ is the analog of the
filtering probability density $P(x_\tau|\delta y_{\textrm{past}})$ and
$\hat E(\delta y_{\textrm{future}}|x_\tau)$ is the analog of the
retrodictive likelihood function $P(\delta
y_{\textrm{future}}|x_\tau)$. One can then solve for the density and
effect operators separately, before combining them to form the
classical smoothing probability density.

\subsection{Filtering}
Since the hybrid density operator can be regarded as a special case of
the density operator, the same tools in quantum measurement theory can
be used to derive a filtering equation for the hybrid operator.
First, write $\hat\rho(x_{t+\delta t}|\delta y_{[t_0,t]})$ in terms of
$\hat\rho(x_t|\delta y_{[t_0,t]})$ as
\begin{align}
\hat\rho(x_{t+\delta t}|\delta y_{[t_0,t]})
&= \int dx_t
\mathcal K(x_{t+\delta t}|x_t)\hat\rho(x_t|\delta y_{[t_0,t]}),
\label{qck}
\end{align}
where $\mathcal K$ is a completely positive map that governs the
Markovian evolution of the hybrid state independent of the measurement
process. Equation (\ref{qck}) may be regarded as a quantum version of
the classical Chapman-Kolmogorov equation.  For infinitesimal $\delta
t$,
\begin{align}
\int dx_t\mathcal K(x_{t+\delta t}|x_t)\hat\rho(x_t)
&\approx \Bk{\bk{\hat 1+\delta t\mathcal L}\hat\rho(x_t=x)}_{x= x_{t+\delta t}}.
\end{align}
The hybrid superoperator $\mathcal L$ can be expressed as
\begin{align}
\mathcal L\hat\rho(x)
&=
\mathcal L_0\hat\rho(x)+\mathcal L_{I}(x)\hat\rho(x)
-
\sum_{\mu}\parti{}{x_\mu}\Bk{A_\mu\hat\rho(x)}
\nonumber\\&\quad
+\frac{1}{2}\sum_{\mu,\nu}\parti{^2}{x_\mu\partial x_\nu}
\Bk{\bk{BQB^T}_{\mu\nu}\hat\rho(x)},
\label{super_L}
\end{align}
where $\mathcal L_0$ governs the evolution of the quantum system,
$\mathcal L_I$ governs the coupling of $x_t$ to the quantum system,
via an interaction Hamiltonian for example, and the last two terms
governs the classical evolution of $x_t$.

Next, write $\hat\rho(x_t|\delta y_{[t_0,t]})$ in terms of
$\hat\rho(x_t|\delta y_{[t_0,t-\delta t]})$ using the quantum Bayes theorem
\cite{gardiner_zoller} as
\begin{align}
\hat\rho(x_t|\delta y_{[t_0,t]})
&=\hat\rho(x_t|\delta y_{[t_0,t-\delta t]},\delta y_t)
\nonumber\\
&=
\frac{\mathcal J(\delta y_t|x_t)\hat\rho(x_t|\delta y_{[t_0,t-\delta t]})}
{\int dx_t\trace(\textrm{numerator})}.
\label{qbayes}
\end{align}
The measurement superoperator $\mathcal J(\delta y_t|x_t)$, a quantum
version of $P(\delta y_t|x_t)$, is defined as
\begin{align}
\mathcal J(\delta y_t|x_t)\hat\rho(x_t)&\equiv
\hat M(\delta y_t|x_t)\hat\rho(x_t)\hat M^\dagger(\delta y_t|x_t).
\end{align}
For infinitesimal $\delta t$ and measurements with Gaussian noise, the
measurement operator $\hat M$ can be approximated as
\cite{diosi}
\begin{align}
\hat M(\delta z_t|x_t)
&\propto
\hat 1 + \sum_{\mu}\gamma_\mu(t)
\bigg[\frac{1}{2}\hat c_\mu(x_t,t)\delta z_{\mu t}
\nonumber\\&\quad
-\frac{\delta t}{8}\hat c^{\dagger}_\mu(x_t,t)\hat c_\mu(x_t,t)\bigg],
\end{align}
where $\delta z_t$ is a vectoral observation process, $\hat c(x_t,t)$
is a vector of hybrid operators, generalized from the purely quantum
$\hat c$ operators in Ref.~\cite{tsang_smooth} so that the
observations may also depend directly on the classical degrees of
freedom, and $\gamma_\mu(t)$ is assumed to be positive.  To cast the
theory in a form similar to the classical one, perform unitary
transformations on $\delta z_t$ and $\hat c$,
\begin{align}
\delta y_t &= U\delta z_t,
\\
\hat C(x_t,t) &= U\hat c(x_t,t),
\end{align}
where $U$ is a unitary matrix, and rewrite the measurement operator as
\begin{align}
\hat M(\delta y_t|x_t)
&\propto
\hat 1 + \frac{1}{2}\hat C^T(x_t,t)R^{-1}(t)\delta y_t
\nonumber\\&\quad
-\frac{\delta t}{8}\hat C^{\dagger T}(x_t,t)R^{-1}(t)\hat C(x_t,t).
\end{align}
$\hat C(x_t,t)$ is a generalization of $C(x_t,t)$ in the classical
case, and $R(t)$ is again a positive-definite matrix that
characterizes the observation uncertainties and is real and symmetric
with eigenvalues $1/\gamma_\mu$. Note that $^\dagger$ is defined as
the adjoint of each vector element, and $^T$ is defined as the matrix
transpose of the vector. For example,
\begin{align}
\hat C^{\dagger T}R^{-1}\hat C \equiv
\sum_{\mu,\nu}\hat C_\mu^{\dagger}(R^{-1})_{\mu\nu}\hat C_\nu.
\end{align}
The evolution of $\hat\rho(x_t|\delta y_{[t_0,t-\delta t]})$ can thus
be calculated by iterating the formula
\begin{align}
&\quad\hat\rho(x_{t+\delta t}|\delta y_{[t_0,t]})
\nonumber\\
&=
\int dx_t \mathcal K(x_{t+\delta t}|x_t)
\frac{\mathcal J(\delta y_t|x_t)\hat\rho(x_t|\delta y_{[t_0,t-\delta t]})}
{\int dx_t\trace(\textrm{numerator})}.
\end{align}
Taking the continuous time limit via It\=o calculus and defining the
conditional hybrid density operator at time $t$ as
\begin{align}
\hat F(x,t) &\equiv \hat\rho(x_t=x|dy_{[t_0,t)}),
\end{align}
one obtains \cite{tsang_smooth}
\begin{align}
d\hat F &= dt
\mathcal L\hat F
\nonumber\\&\quad
+\frac{dt}{8}\bk{2\hat C^TR^{-1}\hat F\hat C^\dagger
-\hat C^{\dagger T}R^{-1}\hat C\hat F -
\hat F \hat C^{\dagger T}R^{-1}\hat C}
\nonumber\\&\quad
+\frac{1}{2}
\Bk{\bk{\hat C-\avg{\hat C}_{\hat F}}^TR^{-1}d\eta_t\hat F
+\textrm{H.c.}},
\label{qks}
\end{align}
where
\begin{align}
\avg{\hat C}_{\hat F} &\equiv \int dx \trace\Bk{\hat C(x,t)\hat F(x,t)},
\\
d\eta_t &\equiv  dy_t - \frac{dt}{2}\avg{\hat C+\hat C^\dagger}_{\hat F}
\end{align}
is a Wiener increment with covariance matrix $R(t)dt$ \cite{diosi},
H.c.\ denotes the Hermitian conjugate, and the initial condition is
the \textit{a priori} hybrid density operator $\hat\rho(x_{t_0})$.
Equation (\ref{qks}) is a quantum version of the KS equation
(\ref{kushner}) and can be regarded as a special case of the Belavkin
quantum filtering equation \cite{belavkin}.

A linear version of the KS equation for an unnormalized $\hat F(x,t)$
is
\begin{align}
d\hat f
&= dt
\mathcal L\hat f
\nonumber\\&\quad
+\frac{dt}{8}\bk{2\hat C^TR^{-1}\hat f\hat C^\dagger
-\hat C^{\dagger T}R^{-1}\hat C\hat f -
\hat f \hat C^{\dagger T}R^{-1}\hat C}
\nonumber\\&\quad
+\frac{1}{2}
\bk{\hat C^TR^{-1}dy_t\hat f
+\textrm{H.c.}},
\label{qzakai}
\end{align}
and the normalization is
\begin{align}
\hat F(x,t) &= \frac{\hat f(x,t)}{\int dx \trace[\hat f(x,t)]}.
\end{align}
Equation (\ref{qzakai}) is a quantum generalization of the DMZ
equation (\ref{zakai}).

\subsection{Retrodiction and smoothing}
Taking a similar approach to the one in Sec.~\ref{retro_smooth} and
using the quantum regression theorem, one can express the future
observation statistics as \cite{carmichael}
\begin{align}
P(\delta y_{\textrm{future}})
&= \int dx_\tau 
\trace\Bk{\hat E(\delta y_{\textrm{future}}|x_\tau)\hat\rho(x_\tau)}
\label{qretro}\\
&=
\int dx_T \trace\bigg[\int dx_{T-\delta t} \mathcal K(x_T|x_{T-\delta t})
\nonumber\\&\quad\cdot
 \mathcal J(\delta y_{T-\delta t}|x_{T-\delta t})
\nonumber\\&\quad\cdot
\int dx_{T-2\delta t}
\mathcal K(x_{T-\delta t}|x_{T-2\delta t})
\nonumber\\&\quad\cdot
 \mathcal J(\delta y_{T-2\delta t}|x_{T-2\delta t})\dots
\nonumber\\&\quad\cdot
\int dx_\tau \mathcal K(x_{\tau+\delta t}|x_\tau)
\mathcal J(\delta y_{\tau}|x_\tau)\hat\rho(x_\tau)\bigg],
\label{qexpand}
\end{align}
which are analogous to Eq.~(\ref{retro}) and Eq.~(\ref{expand}),
respectively. Comparing Eq.~(\ref{qretro}) with Eq.~(\ref{qexpand}),
and defining the adjoint of a superoperator $\mathcal O$
as $\mathcal O^*$, such that
\begin{align}
\trace\Bk{\hat E(x)\mathcal O\hat\rho(x)}
&=
\trace\BK{\Bk{\mathcal O^*\hat E(x)}\hat\rho(x)},
\end{align}
the hybrid effect operator can be written as
\begin{align}
&\quad\hat E(\delta y_{\textrm{future}}|x_\tau)
\nonumber\\
&=\mathcal J^*(\delta y_\tau|x_\tau)
\int dx_{\tau+\delta t}\mathcal K^*(x_{\tau+\delta t}|x_\tau)\dots
\nonumber\\&\quad
\cdot\mathcal J^*(\delta y_{T-2\delta t}|x_{T-2\delta t})
\int dx_{T-\delta t}\mathcal K^*(x_{T-\delta t}|x_{T-2\delta t})
\nonumber\\&\quad
\cdot
\mathcal J^*(\delta y_{T-\delta t}|x_{T-\delta t})
\int dx_T\mathcal K^*(x_T|x_{T-\delta t})\hat 1.
\label{hybrid_effect}
\end{align}
The operation $\mathcal K^* \equiv \int dx'\mathcal K^*(x'|x)\cdot$
may also be regarded as a hybrid superoperator on a hybrid operator,
and is the adjoint of $\mathcal K \equiv \int dx'\mathcal
K(x|x')\cdot$, defined by
\begin{align}
\Avg{\hat E(x),\mathcal K\hat\rho(x)} &= \Avg{\mathcal K^*\hat E(x),\hat\rho(x)},
\end{align}
with respect to the Hilbert-Schmidt inner product
\begin{align}
\Avg{\hat E(x),\hat\rho(x)}
&\equiv \int dx \trace \Bk{\hat E(x)\hat\rho(x)}.
\label{inner}
\end{align}
One can then rewrite Eqs.~(\ref{qretro}), (\ref{qexpand}), and
(\ref{hybrid_effect}) more elegantly as
\begin{align}
\Avg{\hat E(x),\hat\rho(x)}
&= \Avg{\hat 1,\mathcal K\mathcal J\dots\mathcal K\mathcal J\hat\rho(x)},
\\
\hat E(x) &= \mathcal J^*\mathcal K^*\dots\mathcal J^*\mathcal K^*\hat 1.
\end{align}
In the continuous time limit, a linear
stochastic differential equation for the unnormalized
effect operator $\hat g(x,t) \propto \hat E(dy_{[t,T)}|x_t=x)$
can be derived. The result is \cite{tsang_smooth}
\begin{align}
-d\hat g
&= dt
\mathcal L^*\hat g
\nonumber\\&\quad
+\frac{dt}{8}\bk{2\hat C^{\dagger T}R^{-1}\hat g\hat C
-\hat g\hat C^{\dagger T}R^{-1}\hat C -
 \hat C^{\dagger T}R^{-1}\hat C\hat g}
\nonumber\\&\quad
+\frac{1}{2}
\bk{\hat g\hat C^TR^{-1}dy_t
+\textrm{H.c.}},
\label{backward_qzakai}
\end{align}
to be solved backward in time in the backward It\=o sense, with the
final condition
\begin{align}
\hat g(x,t)\propto \hat 1.
\end{align}
Equation (\ref{backward_qzakai}) is the adjoint equation of the
forward quantum DMZ equation (\ref{qzakai}) with respect to the inner
product defined by Eq.~(\ref{inner}). It is a generalization of the
classical backward DMZ equation (\ref{backward_zakai}).

Finally, after solving Eq.~(\ref{qzakai}) for $\hat f(x,\tau)$ and
Eq.~(\ref{backward_qzakai}) for $\hat g(x,\tau)$, the smoothing
probability density is
\begin{align}
h(x,\tau) &\equiv P(x_\tau=x|dy_{[t_0,T)}) 
=
\frac{\trace[\hat g(x,\tau)\hat f(x,\tau)]}
{\int dx\trace[\hat g(x,\tau)\hat f(x,\tau)]}.
\label{qsmooth}
\end{align}
The denominator of Eq.~(\ref{qsmooth}) ensures that $h(x,\tau)$ is
normalized, so $\hat f(x,\tau)$ and $\hat g(x,\tau)$ need not be
normalized separately. Table \ref{analogs} lists some important
quantities in classical smoothing with their generalizations in hybrid
smoothing for comparison.

\begin{table*}[htbp]
\begin{tabular}{|l|p{0.3\textwidth}|l|p{0.3\textwidth}|}
\hline
Classical & Description & Hybrid & Description
\\
\hline
$P(x_{t+\delta t}|x_{t},\delta y_t)$ &
transition probability density, 
appears in Chapman-Kolmogorov equation (\ref{ck}) &
$\mathcal K(x_{t+\delta t}|x_t)$
&
transition superoperator,
appears in quantum Chapman-Kolmogorov equation (\ref{qck})
\\
\hline
$P(\delta y_t|x_t)$ & observation probability density, appears
in Bayes theorem (\ref{bayes2})
&
$\mathcal J(\delta y_t|x_t)$ & measurement superoperator, appears
in quantum Bayes theorem (\ref{qbayes})
\\
\hline
$P(x_t|dy_{[t_0,t)})$, $F(x,t)$ &  filtering probability density,
obeys Kushner-Stratonovich equation (\ref{kushner})  & 
$\hat\rho(x_t|dy_{[t_0,t)})$, $\hat F(x,t)$ & filtering hybrid density operator,
obeys Belavkin equation (\ref{qks})
\\
\hline
$f(x,t)$ & unnormalized $F(x,t)$, obeys Duncan-Mortensen-Zakai (DMZ) equation
(\ref{zakai}) &
$\hat f(x,t)$ & unnormalized $f(x,t)$,
obeys quantum DMZ equation (\ref{qzakai})
\\
\hline
$P(dy_{[t,T)}|x_t)$ & retrodictive likelihood function &
$\hat E(dy_{[t,T)}|x_t)$ & hybrid effect operator
\\
\hline
$g(x,t)$ & unnormalized $P(dy_{[t,T)}|x_t)$, obeys backward DMZ
equation (\ref{backward_zakai}) &
$\hat g(x,t)$ & unnormalized $\hat E(dy_{[t,T)}|x_t)$,
obeys backward quantum DMZ equation (\ref{backward_qzakai})
\\
\hline
$P(x_\tau|dy_{[t_0,T)})$, $h(x,\tau)$ & smoothing probability density,
obeys Eq.~(\ref{smooth}) &
$P(x_\tau|dy_{[t_0,T)})$, $h(x,\tau)$ & smoothing probability density,
obeys Eq.~(\ref{qsmooth})
\\
\hline
\end{tabular}
\caption{Important quantities in classical smoothing and their
  generalizations in hybrid smoothing.}
\label{analogs}
\end{table*}

\subsection{Smoothing in terms of Wigner distributions}
To solve Eqs.~(\ref{qzakai}), (\ref{backward_qzakai}), and
(\ref{qsmooth}), one way is to convert them to equations for
quasiprobability distributions \cite{walls}. The Wigner distribution
is especially useful for quantum systems with continuous degrees of
freedom. It is defined as \cite{walls,mandel}
\begin{align}
f(q,p) &\equiv \frac{1}{2\pi}
\int du \Bra{q-\frac{u}{2}}\hat f\Ket{q+\frac{u}{2}}\exp\bk{ip^Tu},
\end{align}
where $q$ and $p$ are normalized position and momentum vectors.  It
has the desirable property
\begin{align}
\int dq dp g(q,p)f(q,p) &= \frac{1}{2\pi}\trace\bk{\hat g\hat f},
\end{align}
which is unique among generalized quasiprobability distributions
\cite{mandel}.  The smoothing probability density given by
Eq.~(\ref{qsmooth}) can then be rewritten as
\begin{align}
h(x,\tau)
&=
\frac{\int dq dp g(q,p,x,\tau)f(q,p,x,\tau)}
{\int dq dp dx g(q,p,x,\tau)f(q,p,x,\tau)},
\label{wigner_smooth}
\end{align}
where $f(q,p,x,\tau)$ and $g(q,p,x,\tau)$ are the Wigner distributions
of $\hat f$ and $\hat g$, respectively.  Equation
(\ref{wigner_smooth}) resembles the classical expression
(\ref{smooth}) with the quantum degrees of freedom $q$ and $p$
marginalized.  If $f(q,p,x,t_0)$ is nonnegative and the stochastic
equations for $f(q,p,x,t)$ and $g(q,p,x,t)$ converted from
Eqs.~(\ref{qzakai}) and (\ref{backward_qzakai}) have the same form as
the classical DMZ equations given by Eqs.~(\ref{zakai}) and
(\ref{backward_zakai}), the hybrid smoothing problem becomes
equivalent to a classical one and can be solved using well known
classical smoothers.  For example, if $f(q,p,x,t)$ and $g(q,p,x,t)$
are Gaussian, $h(x,\tau)$ is also Gaussian, and their means and
covariances can be solved using the linear MFP smoother described in
Sec.~\ref{mfp}.

\section{\label{adaptive}Phase-locked loop design for narrowband
  squeezed beams}
\begin{figure*}[htbp]
\centerline{\includegraphics[width=0.8\textwidth]{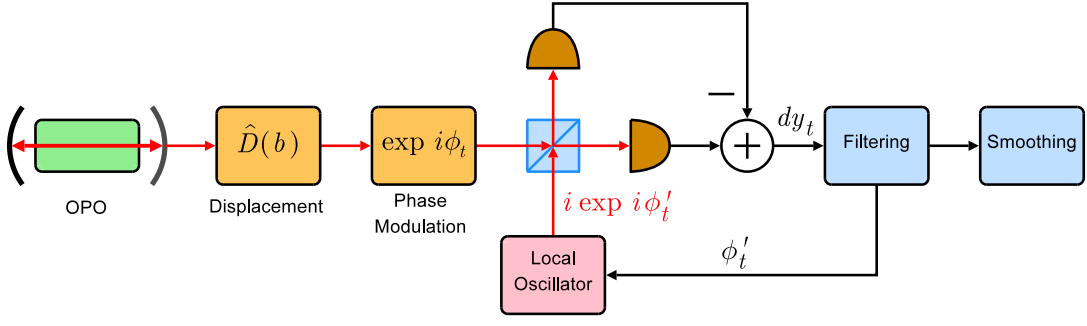}}
\caption{(Color online). Homodyne phase-locked loop (PLL) for phase
  estimation with a narrowband squeezed optical beam produced from an
  optical parametric oscillator (OPO).}
\label{pll}
\end{figure*}

Consider the PLL setup depicted in Fig.~\ref{pll}.  The optical
parametric oscillator (OPO) produces a squeezed vacuum with a squeezed
$p$ quadrature and an antisqueezed $q$ quadrature.  The squeezed
vacuum is then displaced by a real constant $b$ to produce a
phase-squeezed beam, the phase of which is modulated by $\phi_t =
x_{1t}$, an element of the vectoral random process $x_t$ described by
the system It\=o equation (\ref{system}). The output beam is measured
continuously by a homodyne PLL, and the local-oscillator phase
$\phi_t'$ is continuously updated according to the real-time
measurement record.

The use of PLL for phase estimation in the presence of quantum noise
has been mentioned as far back as 1971 by Personick \cite{personick}.
Wiseman suggested an adaptive homodyne scheme to measure a constant
phase \cite{wiseman}, which was then experimentally demonstrated by
Armen \textit{et al.}\ for the optical coherent state
\cite{armen}. Berry and Wiseman \cite{berry1} and Pope \textit{et
  al.}\ \cite{pope} studied the problem with $\phi_t$ being a Wiener
process.  Berry and Wiseman later generalized the theory to account
for narrowband squeezed beams \cite{berry}.  Tsang \textit{et al.}\
also studied the problem for the case of $x_t$ being a Gaussian
process \cite{tsang,tsang2}, but the squeezing model considered in
Refs.~\cite{tsang,tsang2} is not realistic.  Using the hybrid
smoothing theory developed in Sec.~\ref{hybrid}, one can now
generalize these earlier results to the case of an arbitrary diffusive
Markov process and a realistic squeezing model.

Let $\hat\rho(x_t)$ be the hybrid density operator for the combined
quantum-OPO-classical-modulator system. The evolution of the OPO below
threshold in the interaction picture is governed by
\begin{align}
\mathcal L_0\hat\rho(x) &= -\frac{i}{\hbar}\Bk{\hat H_0,\hat\rho(x)},
\\
\hat H_0 &= -i\frac{\hbar\chi}{2} \bk{\hat a\hat a-\hat a^\dagger\hat a^\dagger}
\\
&=\frac{\hbar\chi}{2}\bk{\hat q\hat p + \hat p \hat q},
\end{align}
where $\hat a$ is the annihilation operator for the cavity optical
mode, and $\hat q$ and $\hat p$ are the antisqueezed and squeezed
quadrature operators, respectively, defined as
\begin{align}
\hat q &\equiv \frac{\hat a + \hat a^\dagger}{\sqrt{2}},
\\
\hat p &\equiv \frac{\hat a -\hat a^\dagger}{\sqrt{2}i},
\end{align}
with the commutation relation
\begin{align}
[\hat q,\hat p] &= i.
\end{align}
The classical phase modulator does not influence the
evolution of the OPO, so 
\begin{align}
\mathcal L_I = 0,
\end{align}
but it modulates the OPO output. $\hat C(x_t,t)$ in this case is
\begin{align}
\hat C(x_t,t) &= -2i
\bk{b+\sqrt{\gamma}\hat a }\exp\bk{i\phi_t-i\phi_t'},
\end{align}
where $\gamma$ is the transmission coefficient of the partially
reflecting OPO output mirror, $R = 1$, and the symbol and sign
conventions here roughly follows those of
Refs.~\cite{tsang,tsang2}. To ensure the correct unconditional quantum
dynamics, the Hamiltonian should be changed to
(Ref.~\cite{gardiner_zoller}, Sec.~11.4.3)
\begin{align}
\hat H_0' &= \hat H_0
-i\frac{i\hbar b\sqrt{\gamma} }{2}(\hat a-\hat a^\dagger),
&
\mathcal L_0\hat\rho(x) &= -\frac{i}{\hbar}\Bk{\hat H_0',\hat\rho(x)},
\end{align}
in order to eliminate the spurious effect of the displacement term in
$\hat C$ on the OPO. After some algebra, the forward stochastic
equation for the Wigner distribution $f(q,p,x,t)$ becomes
\begin{align}
df &= dt\bigg\{-
\sum_\mu\parti{}{x_\mu}(A_\mu f)
+\frac{1}{2}\sum_{\mu,\nu}\parti{^2}{x_\mu\partial x_\nu}
\Bk{\bk{BQB^T}_{\mu\nu}f}
\nonumber\\&\quad
-\Bk{\bk{\chi -\frac{\gamma}{2}}\parti{}{q}\bk{
qf}+\bk{-\chi -\frac{\gamma}{2}}\parti{}{p}\bk{pf}}
\nonumber\\&\quad
+\frac{\gamma}{4}\bk{\partit{f}{q}+\partit{f}{p}}\bigg\}
\nonumber\\&\quad
+dy_t
\bigg[\sin(\phi-\phi_t')
\bk{2b+\sqrt{2\gamma}q + \sqrt{\frac{\gamma}{2}}\parti{}{q}}
\nonumber\\&\quad
+\cos(\phi-\phi_t')
\bk{\sqrt{2\gamma}p+\sqrt{\frac{\gamma}{2}}\parti{}{p}}\bigg]f.
\label{wigner_zakai}
\end{align}
This is precisely the classical DMZ equation (\ref{zakai}) with
correlated system and observation noises.  The equivalent classical
system equations are then
\begin{align}
dq_t &= \bk{\chi-\frac{\gamma}{2}}q_t dt
+ \sqrt{\frac{\gamma}{2}} d\alpha_t,
\nonumber\\
dp_t &= \bk{-\chi-\frac{\gamma}{2}}p_t dt
+ \sqrt{\frac{\gamma}{2}} d\beta_t,
\nonumber\\
dx_t &= A(x_t,t) dt + B(x_t,t)dW_t,
\label{equiv_system}
\end{align}
and the equivalent observation equation is
\begin{align}
dy_t &= 
2b \sin(\phi_t-\phi_t') dt
+d\zeta_t,
\nonumber\\
d\zeta_t &\equiv \sin(\phi_t-\phi_t')
\bk{\sqrt{2\gamma} q_tdt-d\alpha_t}
\nonumber\\&\quad
+\cos(\phi_t-\phi_t')\bk{\sqrt{2\gamma}p_tdt -d\beta_t},
\label{equiv_observ}
\end{align}
where $d\alpha_t$ and $d\beta_t$ are independent Wiener increments
with covariance $dt$. $d\alpha_t$ and $d\beta_t$, which appear in both
the system equation and the observation equation, are simply
quadratures of the vacuum field, coupled to both the cavity mode and
the output field via the OPO output mirror.  Equations
(\ref{equiv_system}) and (\ref{equiv_observ}) coincide with the model
of Berry and Wiseman in Ref.~\cite{berry} when $x_t$ is a Wiener
process, and Eq.~(\ref{wigner_zakai}) is the continuous limit of their
approach to phase estimation. This approach can also be regarded as an
example of the general method of accounting for colored observation
noise by modeling the noise as part of the system
\cite{vantrees,crassidis,simon}.

If $\chi = 0$, $d\zeta_t/dt$ is an additive white Gaussian noise, and
the model is reduced to that studied in
Refs.~\cite{berry1,pope,tsang,tsang2}. In that case, it is desirable
to make $\phi_t'$ follow $\phi_t$ as closely as possible, so that
$dy_t$ can be approximated as
\begin{align}
dy_t &\approx 
2b(\phi_t-\phi_t') dt+d\zeta_t,
\label{approx}
\end{align}
and the Kalman filter can be used if $x_t$ is Gaussian
\cite{tsang2}. Provided that Eq.~(\ref{approx}) is valid, one should
make $\phi_t'$ the conditional expectation of $\phi_t=x_{1t}$, given
by
\begin{align}
\phi_t' &= \Avg{\phi_t}_{\hat f} =\int dq dp dx\, x_1 f(q,p,x,t).
\label{lo_phase}
\end{align}
For phase-squeezed beams, it also seems desirable to make $\phi_t'$
close to $\phi_t$ in order to minimize the magnitude of
$d\zeta_t$. Equation (\ref{lo_phase}) may not provide the optimal
$\phi_t'$ in general, however, as it does not necessarily minimize the
magnitude of $d\zeta_t$ or the estimation errors. The optimal control
law for $\phi_t'$ should be studied in the context of control theory.

While $\phi_t'$ needs to be updated in real time and must be
calculated via filtering, the estimation accuracy can be improved by
smoothing.  The backward DMZ equation for $g(q,p,x,t)$ is the adjoint
equation with respect to Eq.~(\ref{wigner_zakai}), given by
\begin{align}
-dg &= dt\bigg\{
\sum_\mu A_\mu\parti{g}{x_\mu}
+\frac{1}{2}\sum_{\mu,\nu}\bk{BQB^T}_{\mu\nu}
\parti{^2g}{x_\mu\partial x_\nu}
\nonumber\\&\quad
+\Bk{\bk{\chi -\frac{\gamma}{2}}q\parti{g}{q}
+\bk{-\chi -\frac{\gamma}{2}}p\parti{g}{p}}
\nonumber\\&\quad
+\frac{\gamma}{4}\bk{\partit{g}{q}+\partit{g}{p}}\bigg\}
\nonumber\\&\quad
+dy_t
\bigg[\sin(\phi-\phi_t')
\bk{2b+\sqrt{2\gamma}q - \sqrt{\frac{\gamma}{2}}\parti{}{q}}
\nonumber\\&\quad
+\cos(\phi-\phi_t')
\bk{\sqrt{2\gamma}p-\sqrt{\frac{\gamma}{2}}\parti{}{p}}\bigg]g,
\label{wigner_bzakai}
\end{align}
and the smoothing probability density $h(x,\tau)$ is given by
Eq.~(\ref{wigner_smooth}). The use of linear smoothing for the case of
$x_t$ being a Gaussian process and $d\zeta_t/dt$ being a white
Gaussian noise has been studied in
Refs.~\cite{tsang,tsang2}. Practical strategies of solving
Eqs.~(\ref{wigner_zakai}) and (\ref{wigner_bzakai}) in general are
beyond the scope of this paper, but classical nonlinear filtering and
smoothing techniques should help
\cite{jazwinski,vantrees,crassidis,simon}.

One can also use the hybrid smoothing theory to study the general
problem of force estimation via a squeezed probe beam and a homodyne
PLL, by modeling the phase modulator as a quantum mechanical
oscillator instead and combining the problem studied in this section
with the force estimation problem studied in Ref.~\cite{tsang_smooth}.

\section{\label{weak}Weak values as quantum smoothing estimates}
\begin{figure}[htbp]
\centerline{\includegraphics[width=0.48\textwidth]{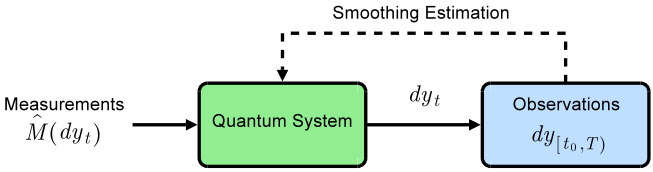}}
\caption{(Color online). The quantum smoothing problem.}
\label{quantum_smoothing}
\end{figure}
Previous sections focus on the estimation of classical signals, but
there is no reason why one cannot apply smoothing to quantum degrees
of freedom as well, as shown in Fig.~\ref{quantum_smoothing}.  First
consider the predicted density operator at time $\tau$ conditioned
upon past observations, given by
\begin{align}
\hat\rho(\tau) &\equiv \frac{\hat f(\tau)}{\trace[\hat f(\tau)]},
\end{align}
where the classical degrees of freedom are neglected for simplicity.
The predicted expectation of an observable, such as the position
of a quantum mechanical oscillator, is
\begin{align}
\avg{\hat O}_{\hat f} &\equiv \trace\Bk{\hat O\hat\rho(\tau)}
=\frac{\trace[\hat O\hat f(\tau)]}{\trace[\hat f(\tau)]}.
\label{predict}
\end{align}
One may also use retrodiction, after some measurements of a quantum
system have been made, to estimate its initial quantum state before
the measurements \cite{barnett,yanagisawa}, using the
retrodictive density operator defined as
\begin{align}
\hat\rho_{\textrm{ret}}(\tau) &\equiv \frac{\hat g(\tau)}{\trace[\hat g(\tau)]}.
\end{align}
The retrodicted expectation of an observable is
\begin{align}
_{\hat g}\avg{\hat O} &\equiv \trace\Bk{\hat\rho_{\textrm{ret}}(\tau)\hat O}
=\frac{\trace[\hat g(\tau)\hat O]}{\trace[\hat g(\tau)]}.
\label{retrodict}
\end{align}
Causality prevents one from going back in time to verify the
retrodicted expectation, but if the degree of freedom with respect to
$\hat O$ at time $\tau$ is entangled with another ``probe'' system,
then one can verify the retrodicted expectation by measuring the probe
and inferring $\hat O$ \cite{yanagisawa}.

The idea of verifying retrodiction by entangling the system at time
$\tau$ with a probe can also be extended to the case of smoothing, as
proposed by Aharonov \textit{et al.}\ \cite{aav}. In the middle of a
sequence of measurements, if one weakly couples the system to a probe
for a short time, so that the system is weakly entangled with the
probe, and the probe is subsequently measured, the measurement outcome
on average can be characterized by the so-called weak value of an
observable, defined as \cite{aav,wiseman_aav}
\begin{align}
_{\hat g}\avg{\hat O}_{\hat f}
&\equiv \frac{\trace[\hat g(\tau)\hat O\hat f(\tau)]}
{\trace[\hat g(\tau)\hat f(\tau)]}.
\end{align}
The weak value becomes a prediction given by Eq.~(\ref{predict}) when
future observations are neglected, such that $\hat g(\tau) = \hat 1$,
and becomes a retrodiction given by Eq.~(\ref{retrodict}) when past
observations are neglected and there is no \textit{a priori}
information about the quantum system at time $\tau$, such that $\hat
f(\tau) = \hat 1$.  When $\hat f(\tau)$ and $\hat g(\tau)$ are
incoherent mixtures of $\hat O$ eigenstates,
\begin{align}
\hat f(\tau) &= \sum_O f(O,\tau)\ket{O}\bra{O},
\\
\hat g(\tau) &= \sum_O g(O,\tau)\ket{O}\bra{O},
\end{align}
the weak value becomes
\begin{align}
_{\hat g}\avg{\hat O}_{\hat f}
&= \frac{\sum_O O g(O,\tau)f(O,\tau)}
{\sum_O g(O,\tau)f(O,\tau)},
\end{align}
and is consistent with the classical time-symmetric smoothing theory
described in Sec.~\ref{classical}. Hence, the weak value can be
regarded as a quantum generalization of the smoothing estimate,
conditioned upon past and future observations. 

One can also establish a correspondence between a classical theory and
a quantum theory via quasiprobability distributions. Given the
smoothing probability density in terms of the Wigner distributions in
Eq.~(\ref{wigner_smooth}), one may be tempted to undo the
marginalizations over the quantum degrees of freedom and define a
smoothing quasiprobability distribution as
\begin{align}
h(q,p,\tau) &= \frac{g(q,p,\tau)f(q,p,\tau)}
{\int dq dp  g(q,p,\tau)f(q,p,\tau)},
\label{smooth_qpd}
\end{align}
where $f(q,p,\tau)$ and $g(q,p,\tau)$ are the Wigner distributions of
$\hat f(\tau)$ and $\hat g(\tau)$, respectively.  Intriguingly,
$h(q,p,\tau)$, being the product of two Wigner distributions, can
exhibit quantum position and momentum uncertainties that violate the
Heisenberg uncertainty principle. This has been shown in
Ref.~\cite{tsang2}, when the position of a quantum mechanical
oscillator is monitored via continuous measurements and smoothing is
applied to the observations.  From the perspective of classical
estimation theory, it is perhaps not surprising that smoothing can
improve upon an uncertainty relation based on a predictive theory.
The important question is whether the sub-Heisenberg uncertainties can
be verified experimentally. Ref.~\cite{tsang2} argues that it can be
done only by Bayesian estimation, but in the following I shall propose
another method based on weak measurements.

It can be shown that the expectation of $q$ using $h(q,p,\tau)$ is
\begin{align}
\avg{q}_h &\equiv \int dq dp \, qh(q,p,\tau)
\\
&=\real\frac{\trace[\hat g(\tau)\hat q\hat f(\tau)]}
{\trace[\hat g(\tau)\hat f(\tau)]}
=\real{} _{\hat g}\avg{\hat q}_{\hat f},
\end{align}
which is the real part of the weak value, and likewise for
$\avg{p}_h$, so the smoothing position and momentum estimates 
are closely related to their weak values.
More generally, consider the joint probability density for a quantum
position measurement followed by a quantum momentum measurement,
conditioned upon past and future observations:
\begin{align}
P(y_q,y_p)
&= 
\frac{1}{\mathcal C}\trace\Bk{\hat g(\tau)\hat M_p(y_p)\hat M_q(y_q)
\hat f(\tau)\hat M_q^\dagger(y_q)\hat M_p^\dagger(y_p)},
\label{joint_measurement}
\\
\mathcal C
&\equiv \int dy_q dy_p
\trace\Big[\hat g(\tau)\hat M_p(y_p)\hat M_q(y_q)
\nonumber\\&\quad\times
\hat f(\tau)\hat M_q^\dagger(y_q)\hat M_p^\dagger(y_p)\Big],
\end{align}
where the measurement operators
\begin{align}
\hat M_q(y_q) &=
\int dq \bk{\frac{\epsilon_q}{2\pi}}^{\frac{1}{4}}
\exp\Bk{-\frac{\epsilon_q}{4}(y_q-q)^2}\ket{q}\bra{q},
\\
\hat M_p(y_p) &=
\int dp \bk{\frac{\epsilon_p}{2\pi}}^{\frac{1}{4}}
\exp\Bk{-\frac{\epsilon_p}{4}(y_p-p)^2}\ket{p}\bra{p}
\end{align}
are assumed to be Gaussian and backaction evading.  After some
algebra,
\begin{align}
P(y_q,y_p) &= \int dq dp
\bk{\frac{\epsilon_q}{2\pi}}^{\frac{1}{2}}
\bk{\frac{\epsilon_p}{2\pi}}^{\frac{1}{2}}
\nonumber\\&\quad\times
\exp\Bk{-\frac{\epsilon_q}{2}(y_q-q)^2-\frac{\epsilon_p}{2}(y_p-p)^2
}\tilde P(q,p),
\label{convolution}\\
\tilde P(q,p) &\equiv 
\frac{1}{2\pi\mathcal C}
\int du dv\exp\bk{-\frac{\epsilon_q u^2+\epsilon_p v^2}{8}}
\nonumber\\&\quad\times
\Bra{p+\frac{v}{2}}\hat g(\tau)\Ket{p-\frac{v}{2}}\exp(ivq)
\nonumber\\&\quad\times
\Bra{q-\frac{u}{2}}\hat f(\tau)\Ket{q+\frac{u}{2}}\exp(ipu).
\end{align}
From the perspective of classical probability theory,
Eq.~(\ref{convolution}) can be interpreted as the probability density
of noisy position and momentum measurements with noise variances
$1/\epsilon_q$ and $1/\epsilon_p$, when the measured object has a
classical phase-space density given by $\tilde P(q,p)$. In the limit
of infinitesimally weak measurements, $\epsilon_q, \epsilon_p \to 0$,
and
\begin{align}
\lim_{\epsilon_q,\epsilon_p\to 0}\tilde P(q,p) &= h(q,p,\tau).
\end{align}
Thus, $h(q,p,\tau)$ can be obtained approximately from an experiment
with small $\epsilon_q$ and $\epsilon_p$ by measuring $P(y_q,y_p)$ for
the same $\hat g$ and $\hat f$ and deconvolving
Eq.~(\ref{convolution}).  In practice, $\epsilon_q$ and $\epsilon_p$
only need to be small enough such that $\tilde P(q,p)\approx
h(q,p,\tau)$.  This allows one, at least in principle, to
experimentally demonstrate the sub-Heisenberg uncertainties predicted
in Ref.~\cite{tsang2} in a frequentist way, not just by Bayesian
estimation as described in Ref.~\cite{tsang2}.  Note, however, that
$h(q,p,\tau)$ can still go negative, so it cannot always be regarded
as a classical probability density. This underlines the wave nature of
a quantum object and may be related to the negative probabilities
encountered in the use of weak values to explain Hardy's paradox
\cite{aav_hardy}.

\section{\label{conclusion}Conclusion}
In conclusion, I have used a discrete-time approach to derive the
classical and quantum theories of time-symmetric smoothing. The hybrid
smoothing theory is applied to the design of PLL, and the relation
between the proposed theory and Aharonov \textit{et al.}'s weak value
theory is discussed. Possible generalizations of the theory include
taking jumps into account for the classical random process
\cite{gardiner} and adding quantum measurements with Poisson
statistics, such as photon counting
\cite{gardiner_zoller,carmichael,walls,mandel}. Potential applications
not discussed in this paper include cavity quantum electrodynamics
\cite{gardiner_zoller,carmichael,walls,mandel}, photodetection theory
\cite{was,gardiner_zoller,mandel}, atomic magnetometry \cite{budker},
and quantum information processing in general.  On a more fundamental
level, it might also be interesting to generalize the weak value
theory and the smoothing quasiprobability distribution to other kinds
of quantum degrees of freedom in addition to position and momentum,
such as spin, photon number, and phase. A general quantum smoothing
theory would complete the correspondence between classical and quantum
estimation theories.

\section*{Acknowledgments}
Discussions with Seth Lloyd and Jeffrey Shapiro are gratefully
acknowledged. This work is financially supported by the Keck
Foundation Center for Extreme Quantum Information Theory.

\end{document}